\documentclass[aps,prl,reprint,twocolumn,superscriptaddress,amsmath,amssymb,citeautoscript,bibnotes]{revtex4-1}
\usepackage{amsmath}
\usepackage{amssymb}
\usepackage{graphicx}
\usepackage{color}
\usepackage{bm}
\usepackage{epstopdf}
\usepackage{grffile}

\DeclareGraphicsExtensions{.eps}

\newcommand{\e}{\rm{e}}
\newcommand{\ii}{\rm{i}}

\newcommand{\EllipticF}{\mathcal{F}}

\begin{document}

\title{Vortex-hole duality: a unified picture of weak and strong-coupling regimes of bosonic ladders with flux}
\author{S. Greschner}
\affiliation{Institut f\"ur Theoretische Physik, Leibniz Universit\"at Hannover, 30167~Hannover, Germany} 
\author{T. Vekua}
\affiliation{James Franck Institute, The University of Chicago, Chicago IL 60637, USA} 
\begin{abstract}
Two-leg bosonic ladders with flux harbor a remarkable vortex-hole duality between the weak-coupling vortex lattice superfluids and strong-coupling charge-density-wave crystals. The strong-coupling crystalline states, which are realized in the vicinity of $\pi$-flux, are independent of particle statistics, and are related with the incompressible fractional quantum Hall states in the thin-cylinder limit. These fully gapped ground states, away of $\pi$-flux, develop nonzero chiral (spin) currents. Contact-interacting quantum gases permit exploration of this vortex-hole duality in experiments.
\end{abstract}
\date{\today}
\maketitle


{\it Dualites} encode important non-perturbative information in statistical, condensed matter and high-energy physics, by mapping weak and strong coupling regimes and providing a way for their unified description~\cite{Seiberg2016}.

A quantum system, depending on conditions, can manifest one of its dual natures profoundly. In a weakly coupled gas or liquid, where positions of particles are not fixed, at sufficiently low temperatures quantum effects set in, and, as a result, Bose particles can develop phase coherence and superfluidity. For strong repulsive inter-particle interactions, crystals can form, where each particle is localized to a certain position in space to get as far as possible from the others. Phases of particles, being conjugate variables of densities, fluctuate strongly in crystals. 
Fluids can develop eddy currents, or vortices when excited. In superfluids with global phase coherence, vortices get topological protection by quantization. Crystals also harbour excitations of topological nature - e.g. point defects such as vacancies (holes).

\begin{figure}[tb]
\centering
\includegraphics[width=0.9\columnwidth]{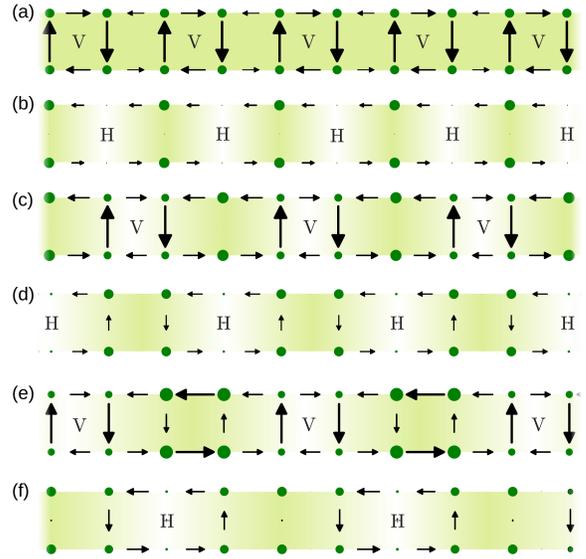}
\caption{Microscopic structures of vortex-hole dual configurations of weak (a), (c), (e) and strong-coupling (b), (d), (f) ground states of bosonic ladders with flux. Dual configurations are (a) $\rho_V=1/2$ vs (b) $\rho_H=1/2$, (c) $\rho_V=1/3$ vs (d)  $\rho_H=1/3$ and (e) $\rho_V=1/4$ vs (f)  $\rho_H=1/4$. Note, that in (a) particle densities are uniform along the ladder, and in (b) particle currents do not show modulations. In contrast, in (c) and (e) particle densities show modulations, similar to particle currents in (d) and (f). }
\label{fig:patterns}
\end{figure}

The purpose of this letter is to demonstrate a spectacular duality between the topological defects of superfluids and crystals, a vortex-hole duality, realized between weak and strong-coupling regimes of bosonic ladders with flux.

Fig.~\ref{fig:patterns} shows the microscopic configurations of local particle currents (arrows) and densities (filled circles) of a few dual weak and strong-coupling ground states of bosonic ladders with flux. In weak-coupling limit the phases of particles are the relevant degrees of freedom, whereas in strong-coupling particle densities play a dominant role. Vortices are indicated by letter V in those plaquettes of Fig.~\ref{fig:patterns}, where $\int_\Box \nabla \Theta dl=2\pi+\phi$, where $\Theta$ is local phase and integration is along the boundary $l$ of the plaquette $\Box$. Holes, defects of the local particle density distribution, are localized on rungs, indicated by letter H. Vortices (elementary loop-currents), topological excitations of weak-coupling regime, repel each other~\cite{Kardar1986} (like same pole magnets) and vortex lattices (VL) at commensurate vortex density $\rho_V$ are dual to hole crystals of charge-density-wave (CDW) states at $\rho_H=\rho_V$ realized in strong-coupling regime, as we will show. Table~\ref{tab:dualities} summarizes the weak and strong-coupling duality relations.
In the weak coupling regime of bosonic ladders few VL superfluids were observed \cite{Greschner2015, Greschner2016} to survive quantum fluctuations on top of classical Josephson-junction (JJ) limit~\cite{Orignac2001}. 
A vortex in classical JJ limit, where phase at each ladder site has definite value, carries a quantum of a fluxoid and is localized on $\xi_V\sim \sqrt{J/2J_\perp}$ plaquettes~\cite{Kardar1986}. Numerical simulations of Bose-Hubbard model on a two-leg ladder with flux showed that particle densities get depleted in the plaquettes where vortices sit, when including quantum fluctuations, and topological excitations of the VL states are domain walls, carrying fractional fluxoids \cite{Greschner2015, Greschner2016}. In this work we will study the strong-coupling ground states of bosonic ladders with flux.

{\it Contact-interacting} cold quantum gases loaded in one-dimensional lattices, with additional second 'synthetic' dimension, can explore this duality in the presence of a homogeneous gauge field. The quantum engineering of synthetic orbital magnetism in neutral cold atom optical-lattices has achieved a tremendous progress during the recent years~\cite{Aidelsburger2013,Miyake2013,Atala2014}. In particular the synthetic-dimension approach~\cite{Celi2014}, that combines a one dimensional optical lattice system with laser assisted transitions between the $M$ internal degrees of freedom which form a compact artificial rung-dimension, allowed for further promising experimental realizations of $M$-leg ladder-like lattices with an artificial magnetic flux~\cite{Stuhl2015,Mancini2015,Livi2016}. 
Since all particles on the same synthetic dimensional rung share the same optical lattice site, contact-interactions lead to exotic long-ranged interactions along the rungs, which for typical systems~\cite{Stuhl2015,Mancini2015} may be assumed to be $SU(M)$ symmetric. The interplay of long-ranged interactions along the synthetic dimension and homogeneous gauge fields has attracted a considerable recent attention, as it gives rise to the ground states bearing analogies with quantum Hall-like behavior~\cite{Barbarino2015,Zeng2015,Taddia2016,Petrescu2015,Cornfeld2015,Petrescu2016,Strinati2016}, or exhibiting exotic quantum magnetism~\cite{Ghosh2015,Ghosh2016,Roux2007,Carr2006,Yan2015,Piraud2015,DiDio2015,Kolley2015,Natu2015,Petrescu2015,Anisimovas2016,Greschner2016}.

\begin{table}
\begin{tabular}{cccl}
 Weak-coupling (JJ) & \hspace{-2.7cm }\vline &  \hspace{-2.5cm}Strong-coupling (quantum Hall)\\
\hline
 Particle phases, flux & \hspace{-2.8cm} $\leftarrow\!$\vline$\!\rightarrow$ &   \hspace{-2.8cm} Particle densities, chem. potential\\
\hline
 Meissner state & \hspace{-2.7cm}$\leftarrow\!$\vline$\!\rightarrow$ & \hspace{-3.5cm}Mott insulator \\
& \hspace{-2.0cm}Topological excitations &\\
 Vortices &\hspace{-2.8cm} $\leftarrow\!$\vline$\!\rightarrow$ & \hspace{-3.6cm}Holes \\
\hline
 Vortex lattices, $\rho_V$ & \hspace{-2.8cm} $\leftarrow\!$\vline$\!\rightarrow$ & \hspace{-2.9cm}Charge-Density-Waves, $\rho_H=\rho_V$ \\
&\hspace{-1.4cm}Top. excitations (domain walls)&\\
Fractional fluxoids &\hspace{-2.7cm}$\leftarrow\!$\vline$\!\rightarrow$ & \hspace{-3cm} Fractional charge\\
\hline
Vortex liquids &\hspace{-2.7cm}$\leftarrow\!$\vline$\!\rightarrow$ & \hspace{-3cm} Superfluids\\
\hline
\end{tabular}
\caption{\label{tab:duality}Duality relations between weak and strong-coupling regimes of bosonic ladders with flux. VL states shown in Fig. 1 (a), (c), and (e) survive moderate quantum fluctuations, due to the coherence of the multi-boson tunnelings between the ladder legs.
}
\label{tab:dualities}
\end{table}

{\it Model-} We consider two-component, $M=2$, case and introduce index $\zeta=0,1$, running along the synthetic dimension. Our microscopic model is a one-dimensional $SU(2)$ symmetric Bose-Hubbard model with spin-orbit coupling, which is equivalent to spinless bosons on two-leg ladder with flux and with the same onsite interactions as interactions along rungs,

\begin{align}
\label{eq:H}
H =& - J \sum_{j=1; \zeta=0,1}^L [b_{j+1,\zeta}^\dagger b_{j,\zeta} +b^\dagger_{j,\zeta}b_{j+1,\zeta}]-\mu\sum_{j}n_j \\ 
&-  J_\perp \sum_{j} [ \e^{\ii \phi j} b_{j,1}^\dagger b_{j,0} + \rm{H.c.}] + \frac{U}{2} \sum_{j,\zeta,\zeta'} n_{j,\zeta} n_{j,\zeta'}\nonumber.
\end{align}
$b_{j,\zeta}$ denotes bosonic annihilation operator on ladder site $j,\zeta$ and $n_j=\sum_{\zeta}n_{j,\zeta}$ denotes particle density on rung $j$. $L$ is the total number of sites along the real space direction, hoppings along ladder legs/rungs are denoted by $J$/$J_\perp$ respectively and $U$ is the Hubbard interaction strength.

We will study Model~\eqref{eq:H} for $U\gg J,J_\perp$, which is relevant for experiments involving a confinement by a deep optical lattice where the interaction $U$ becomes the dominant energy scale. We consider the limit of $U\to\infty$, the so called rung-hard-core limit and address the effects of finite $U$ in supplementary materials~\cite{supmat}\nocite{Granato1990, Mazo1995, Denniston1995, Griffiths1986, Sun2015, Vekua2016, Holzhey1994, Vidal2003, Korepin2004, Calabrese2004}. Since the exchange of particles is forbidden in the rung-hard-core limit, we need not specify statistics of the particles. Particle density is denoted by $\rho= \sum_j \langle n_j \rangle/L=N/L$. In particular in the rung-hard-core limit the maximal particle density is one particle per ladder rung $\rho=1$. The density of holes is defined as $\rho_H=1-\rho$.

One immediately notices that for a $\pi$-flux, the spiraling in-plane magnetic field of the Model~\eqref{eq:H} becomes a staggered field directed along $x$ axes in spin space. Hence we define an order parameter, corresponding to emergent $U(1)$ symmetry at $\pi$-flux - the expectation value of $2S^x=\sum_j 2S_j^x= \sum_j b^{\dagger}_{j,0}b_{j,1}+H.c.$. 

{\it CDW states -} Remarkably, as we will show, the staggered field hard-core Hubbard model exhibits a devil-staircase like structure of CDW phases at fractional fillings $1/2\le \rho<1$. These CDW states are stabilized due to the effective interplay between interactions and strong magnetic field which tends to localize the particle in tight cyclotron orbitals. At unit-filling $\rho=1$, the ground state is a perfect N\'eel-Mott insulator state $ 2\langle S_j^x\rangle=(-1)^j$  (at $U=\infty$, $\pi$-flux and $\rho=1$ the N\'eel-Mott insulator is an exact eigenstate and ground state of Model~\eqref{eq:H}). The staggered field induced N\'eel order at $\phi=\pi$ plays a crucial role in localizing the holes and for emergence of CDW states. This becomes clear if one introduces a single hole on top of unit-filling for $U=\infty$. Then, since particles can not pass each other and since hopping does not flip spin of the particles, hole motion will scramble N\'eel order, by creating a string of displaced particles, which tends to bind the single hole to their initial positions~\cite{supmat}.

Nevertheless, one can imagine that two nearest-neighbour holes can move together by forming a bound state, avoiding frustration of the N\'eel order. The analytic solution of two-hole problem for Fermi-Hubbard model shows that two holes when introduced on top of the N\'eel-Mott state, form bound states (and this happens in fourth order of $J$) only for $U<U_c$, where $U_c\simeq 4J^{\frac{5}{3}}/J_\perp^{\frac{2}{3}}$ for $J_\perp \ll J$ and $U_c=4\sqrt{6}J_\perp$ for $J_\perp \gg J$ \cite{Jaklic1993}. 
At $U=\infty$ (where Fermi and Bose-Hubbard models are equivalent), in the ground state holes stay far apart of each other \cite{supmat}, and due to localized character of single-hole states the CDW phases are formed at rational (commensurate with lattice) hole densities, exactly in the same way as VL states are formed in classical JJ limit~\cite{Kardar1986}: it is a result of the competition between the repulsion among holes (that tries to space holes uniformly apart of each other) and $J_\perp$ that binds holes to rungs.

It is also expected from the above considerations at $\pi$-flux (especially in the limit of tightly localized on-rung holes $J_\perp \gg J$) that the largest hole density for CDW states in the case of contact interactions exhibits a hole on every other rung, the CDW state at $\rho_H=\rho=1/2$, which is also a fully polarized state. 
When adding holes on top of the CDW state at $\rho_H=1/2$, they can move via second order processes $\sim J^2/J_\perp$
maintaining the fully polarized background. Hence states for $\rho_H>1/2$ are expected to be gapless, but showing CDW order with a periodicity of 2 rungs, referred as supersolid for $J_\perp \gg J$~\cite{Bilitewski2016}. 
One can easily obtain analytical ground state properties of this supersolid ground state for any $J_\perp$ in hard-core limit at $\pi$-flux. Hence, for $0<\rho<1/2$ we obtain for the equation of state $\rho(\mu) = \arccos\left[\frac{-J_\perp^2+\mu^2}{2 J^2}-1\right] /\pi$. 
The spontaneously developed density imbalance between the even and odd rungs in the supersolid ground state, $\mathcal{O}_{CDW} = \sum_j (-)^j \langle n_j\rangle / L$, is for any $J_\perp$ given by,
$
\mathcal{O}_{CDW} = \frac{J_\perp}{\pi \sqrt{4 J^2+J_\perp^2} } \, \EllipticF\left[\pi \rho,\frac{4 J^2}{4 J^2+J_\perp^2}\right],
$
with the elliptic integral of the second kind $\EllipticF(\phi,k)$. Note, that CDW order of the supersolid state saturates to a maximal possible value for $J_\perp/J\to\infty$ \cite{Bilitewski2016}.  

\begin{figure}[tb]
\centering
\includegraphics[width=0.45\columnwidth]{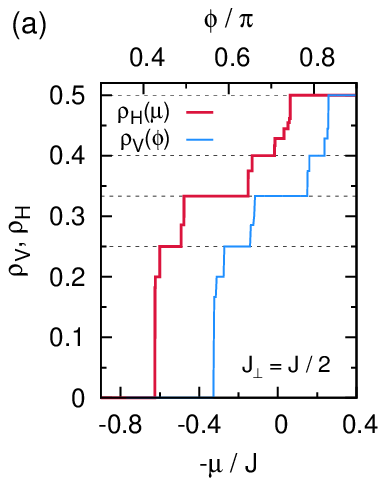}
\includegraphics[width=0.45\columnwidth]{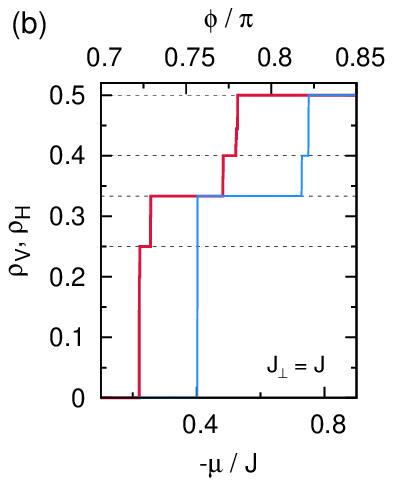}
\caption{The dependence of vortex density on flux $\rho_V(\phi)$ in weak-coupling classical JJ limit and the hole density dependence on chemical potential $\rho_H(\mu)$ in the strong-coupling $U=\infty$, $\phi=\pi$ limit~\cite{supmat} due to vortex-hole duality exhibit a remarkable similarity. With doubling $J_\perp/J$ from (a) to (b) fewer plateaus are formed in $\rho_V(\phi)$ and $\rho_H(\mu)$ curves. 
}
\label{fig:eqs_comp}
\end{figure}

For finite $U<\infty$ at $\phi=\pi$, a single hole on top of unit-filling can gain kinetic energy by second-order hopping $\sim J^2/(U+2J_\perp)$ without leaving behind a perturbed N\'eel string, hence CDW states with $\rho_H \ll 1$ get washed-out quickly for $U<\infty$. In addition, with reducing $U$ from hard-core, particle statistics starts to show up and CDW and supersolid states turn out to be more robust for bosons than for fermions~\cite{BosonFerroNote}. A detailed numerical analysis~\cite{supmat} shows that for example the CDW$_{2/3}$ phase remains stable for $U\gtrsim 20J$ for bosons and $U\gtrsim 30J$ for fermions.

\begin{figure}[tb]
\centering
\includegraphics[width=1.0\columnwidth]{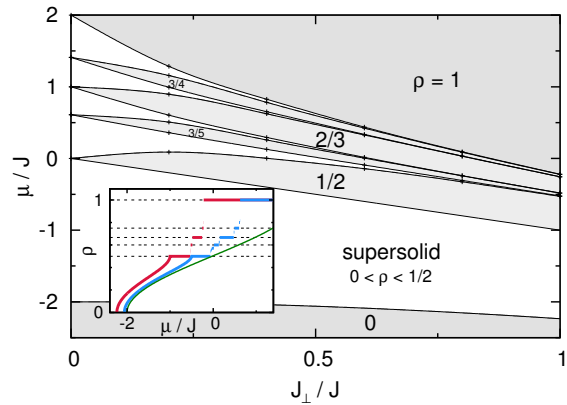}
\caption{Ground state phase diagram in the parameter space of $\mu$  and $J_\perp$ for $U=\infty$ and $\phi=\pi$. For clarity only the most stable fractional CDW phases are shown. Inset shows equation of state $\rho=\rho(\mu)$ of the $U=\infty$ ladder at $\pi$-flux for (from left to right) $J_\perp=J$, $J/2$ and $0$. }
\label{fig:pd_pi}
\end{figure}

{\it Numerics -} In order to obtain a quantitative phase diagram in strong-coupling we perform density matrix renormalization group (DMRG)~\cite{White1992, Schollwoeck2011} calculations of Model~\eqref{eq:H}. For $\rho\geq 1/2$ and $\phi=\pi$, due to strong localization of the holes, infinite DMRG simulations turn out to be very efficient and give extent and structure of the (largest) CDW-phases consistent with the results of finite system size DMRG-simulations~\cite{supmat}. 

Fig.~\ref{fig:patterns} compares dual configurations of local particle densities and currents of weak and strong coupling regimes of bosonic ladders. The microscopic structures of weak-coupling configurations have already been obtained for a single-component Bose-Hubbard model on a two-leg ladder with flux in Ref.~\cite{Greschner2015,Greschner2016}. Strong-coupling configurations are obtained slightly away of $\pi$-flux, for Model (1) at $U=\infty$, where one can see that for $\rho_H<1/2$ local rung and leg-currents also show modulations. Exactly at $\pi$-flux all currents vanish in strong-coupling limit. However, away of $\pi-$flux CDW states support non-zero chiral (spin) current.

In Fig.~\ref{fig:eqs_comp}, for different values of $J_\perp/J$, we compare the vortex-density-vs-flux curves, obtained from phase only model corresponding to classical JJ limit of bosonic ladder \cite{supmat}, with the hole-density-vs-chemical-potential curves obtained for Model (1) at $U=\infty$ at $\pi$-flux.

The phase diagram at $\pi$-flux for $U=\infty$, in the parameter space of $\mu/J$ and $J_\perp/J$, is summarized in Fig.~\ref{fig:pd_pi}. The inset shows $\rho(\mu)$-curves for different values of $J_\perp/J$. 

\begin{figure}[tb]
\centering
\includegraphics[width=1.0\columnwidth]{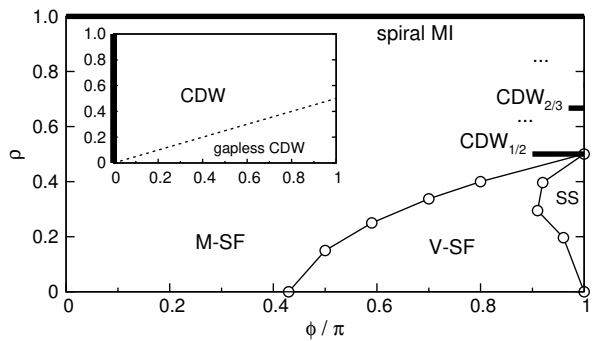}
\caption{Ground state phases in parameter space of flux and density for $U=\infty$ and $J_\perp=J$. Dots indicate CDW states at rational values of $\rho>1/2$ at $\phi=\pi$~\cite{LaughlinPrecursornote}. Inset shows corresponding phase diagram for $J\ll J_\perp$ when number of legs $M\to \infty$.}
\label{fig:pd_phi}
\end{figure}

In the weak-coupling picture including quantum fluctuations (of phases) introduces a mobility of vortices, that can melt VL states into vortex liquids~\cite{Orignac2001}. Analogously in the strong-coupling limit away of $\pi$-flux CDW crystals can melt into superfluids~\cite{supmat}. In Fig.~\ref{fig:pd_phi} we present the ground state phase diagram as function of $\phi$ and $\rho$ for $U=\infty$. Besides Meissner (M-SF) and vortex superfluid (V-SF) phases we observe a Meissner Mott-insulator (spiral MI) phase (which at $\phi=\pi$ evolves into N\'eel MI) and for $\phi \simeq \pi$, the emerging devil's staircase like set of CDW phases at fractional fillings and a supersolid (SS) ground state~\cite{FlatbandBosonizationnote}.

{\it Relation with quantum Hall - } In the following we will consider cases corresponding to $M>2$ and assume that $L\gg M$. We note that region of stability of the CDW-phases may be considerably increased with increasing number of components $M$ and assuming periodic boundary conditions along rungs (cylinder or torus geometry). Moreover, following the discussion of Ref.~\cite{Barbarino2015}, increasing $M$ allows us to relate observed CDW states to fractional quantum Hall states in thin-cylinder limit. In case of $M>2$-leg cylinder, the Hoffstadter-Harper model Eq.~\eqref{eq:H} is consistently defined for a flux $\phi$ multiple of $\frac{2\pi}{M}$. 
We have checked that up to $M=6$ at $\frac{2\pi}{M}$ flux, similar picture as described for two-leg ladder at $\pi$-flux holds. Namely, for low densities, $\rho<\frac{1}{M}$, ground states are gapless, with spontaneously formed long-range modulated density, CDW$_{\frac{1}{M}}$. For densities $\rho>\frac{1}{M}$ a devil's staircase like set of CDW states is expected to emerge. For $U=\infty$ limit same arguments can be used to explain this picture, as for the two-leg ladders at $\pi$-flux~\cite{supmat}. This leads us in the limit $M\to \infty$ to the ground state phase diagram as sketched in the inset of Fig.~\ref{fig:pd_phi}. For particle fillings $\rho<\frac{\phi}{2\pi}$ the ground states are gapless and exhibit a CDW-order of period $\frac{2\pi}{\phi}$. For larger fillings $\rho>\frac{\phi}{2\pi}$ a region of fully gapped CDW$_\rho$ states is formed due to the interplay of commensurate density and periodic potential.

The above discussion allows us to follow the relation between the emerging devil's staircase of fractional CDW-phases for $M$-leg cylinder at $\phi=2\pi/M$, realized for $1/M \le \rho \le 1$ with the similar incompressible states of quantum Hall system in thin-cylinder limit~\cite{Tao1983, Seidel2005,Bergholtz2007, Rotondo2016, Saito2016} realized for filling $ 1/M\le \nu\le 1 $~\cite{supmat}. Recent works~\cite{Barbarino2015, Zeng2015, Taddia2016} indicate that CDW states of $M$-leg ladders approach corresponding fractional quantum Hall states also in topological properties with increasing $M$.

{\it Summarizing,} we have presented a unifying view of weak and strong-coupling physics of interacting bosons on two-leg ladders with flux based on vortex-hole duality. This is a broader version of exact duality mappings (such as Kramers-Wannier duality) and implies the equivalence between: the mechanisms of the emergence of VL and CDW states, the ground state degeneracies of the dual configurations, and quantum numbers of topological excitations on top of the dual ground states. All these properties of weak and strong-coupling dual ground states are identical under vortex-hole exchange. Distinguishing properties of weak-coupling ground states from their dual strong-coupling counterparts is that the former are gapless superfluids, while the latter are gapped crystals.

We also showed that strong contact interactions give rise to a rich phase diagram for (fermionic as well as bosonic) quantum gases in a one-dimensional lattice, with an additional second synthetic dimension, in the presence of the uniform gauge field. In particular devil's staircase-like structure of CDW states emerges at $\pi$-flux without the need of long-range (e.g. dipolar) interactions between the particles, which can be related with the fractional quantum Hall states in thin-cylinder limit~\cite{Zeng2015, Barbarino2015, Saito2016}. Hence, two cornerstone condensed matter systems (both defined on two-leg ladder lattices) - the classical Josephson-junctions array and the quantum Hall system - can be related to each other through the vortex-hole duality~\cite{dualityJJnote}\nocite{Fisher1990,Fazio1991,Blanter1997}. 

On practical side, due to duality, we expect that in thermodynamic limit a critical value of hopping anisotropy exists in strong coupling limit $J^c_\perp/J$, like Aubry's breaking-of-analyticity point in weak-coupling classical JJ limit \cite{Aubry1982}, where devil's staircase of density vs chemical potential curve changes from incomplete to complete one. This expectation is consistent with our numerical observation shown in Fig.~\ref{fig:eqs_comp} and in the inset of Fig.~\ref{fig:pd_pi}, where one can see that with increasing $J_{\perp}/J$ less and less densities are realized in the CDW staircase as function of the chemical potential.

\begin{acknowledgments}
We are grateful to F. Heidrich-Meisner, D. Pirtskhalava, C. Ortix, T. Giamarchi, L. Santos, and P. Wiegmann for useful discussions. S.G. acknowledges support of the German Research Foundation DFG (project no. SA 1031/10-1). TV was supported in part by the National Science Foundation under the Grants NSF DMR-1206648. Simulations were carried out on the cluster system at the Leibniz University of Hannover, Germany.
\end{acknowledgments}

\bibliography{references}

\end{document}


\title{Supplemental Materials to ''Vortex-hole duality: a unified picture of weak and strong-coupling regimes of bosonic ladders with flux``}
\author{S. Greschner}
\affiliation{Institut f\"ur Theoretische Physik, Leibniz Universit\"at Hannover, 30167~Hannover, Germany} 
\author{T. Vekua}
\affiliation{James Franck Institute, The University of Chicago, Chicago IL 60637, USA} 
\begin{abstract}
In the supplementary materials we provide details on the localization of single and paired holes for Model~(1) of the main text for $U=\infty$ and $\phi=\pi$, and discuss numerical methods employed in this paper. We show examples for the stability of the observed ground-state phases for $\phi<\pi$ and $U=\infty$ and for $U<\infty$ and $\phi=\pi$ (separately for bosons and for fermions). Furthermore, we provide an additional discussion on the connection of the CDW states of the $M$-leg ladder model with the incompressible states of the fractional quantum Hall effect in thin-cylinder limit.
\end{abstract}
\date{\today}
\maketitle

\subsection{Localization of a single and a pair of holes}

\begin{figure}[b]
\centering
\includegraphics[width = 1\linewidth]{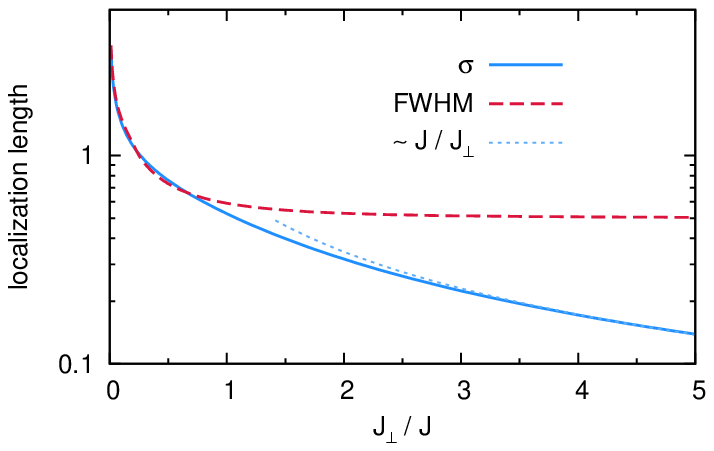}
\caption{Localization length of a single hole (in units of lattice constant) for Model~(S1), as function of $J_\perp/J$. FWHM stands for full width at half maximum and $\sigma$ is a standard deviation. For small $J_\perp/J$ wave function is localized as $\sim e^{-a |j|^b}$, with $a^{-1}\simeq b \simeq 1.5$ and FWHM $\sim (J/J_\perp)^{\alpha}$, with $\alpha \simeq 1/3$. For $J_\perp \gtrsim J/4$ we find that the wave function is localized within three lattice sites with $\sigma<1$ and $\sigma\sim J/J_\perp$ for $J_\perp \gtrsim J$.}
\label{fig:1hole}
\end{figure}

\begin{figure}[b]
\centering
\includegraphics[width = 1\linewidth]{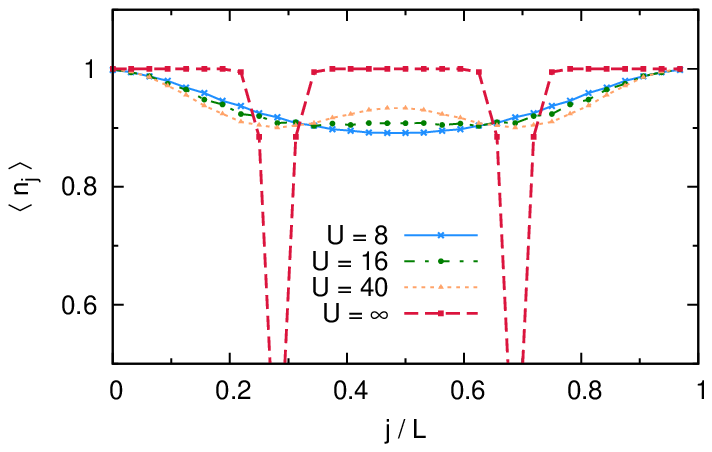}
\caption{Local particle density in the ground state $\langle n_j\rangle$. DMRG-simulation for open boundary conditions of the Model~(1), introduced in the main text, with $L=32$ rungs and $N=30$ (fermionic) particles for $J_\perp/J=1$, $\phi=\pi$ and different values of the interactions strength $U$ (in units of $J$).}
\label{fig:2holes}
\end{figure}

As discussed in the main text we identify the localization of holes introduced on top of the (perfect) N\'eel ordered state at unit-filling, together with the repulsion between the holes for strong $U$, as an underlying mechanism of the formation of crystalline phases at commensurate hole-fillings for $1/2\le \rho<1$ and $\phi=\pi$. The motion of a single hole on top of unit-filling for $U=\infty$ will upset N\'eel order, by creating a string of displaced particles. Hence, a linear confining potential $V_j=2J_\perp|j|$ will act on a single hole introduced at $j=0$, which will be governed by the Hamiltonian 
\begin{equation}
\label{HSH}
H_{d} =- \sum_{j} \{ J ( d_{j+1}^\dagger d_{j}+H.c.)+2J_\perp|j|d_{j}^\dagger d_{j} \}.
\end{equation}
From the Hamiltonian (\ref{HSH}) it is evident that a single-hole state will be localized at the place where it was introduced, and we estimate typical extent of a single hole wavefunction (hole localization length) numerically as shown in Fig. \ref{fig:1hole}.


If the confining potential $V_j$ changes weakly over many sites, for $J_\perp/J\to 0$, one can assume a continuous limit $j \to x$, to obtain the Schr\"odinger equation for a sinlge hole wavefunction $\Psi$, $-J\frac{d^2}{dx^2}\Psi +2J_\perp |x| \Psi=E\Psi$. Since the potential $\sim|x|$ is even, the ground-state solution must be even and hence must have zero derivative at the origin, $\Psi_0(x)=\Psi_0(-x)$, $\Psi_0^{'}(0)=0$. It is given by Airy function of the first kind and allows to estimate localization lengh of a hole as $\xi_H \sim (J/J_\perp)^{1/3}$. This agrees with our numerical simulations shown in Fig. \ref{fig:1hole} for $J_\perp/ J\ll 1$ and with the analogous result obtained for the $t-J$ model \cite{Jaklic1993}.

In Fig.~\ref{fig:2holes} we numerically study the ground-state local particle density of Model~(1) of the main text, for the doping of two holes on top of unit-filling. While for small values of $U/J$ the two holes tend to stay next to each other (signature of the bound state) and two-hole state looks extended (decreasing at the boundary rungs of the ladder), in the rung-hard-core limit the two holes are clearly localized apart of each other (and stay away of boundaries).

\subsection{Numerical methods}

Quantitative results of the strong- (and moderate, as shown below) coupling regime are obtained by means of DMRG simulations of Model~(1) of the main text. Generally we study systems up to $L=180$ rungs keeping typically $D=1000$ matrix-states for open-boundary conditions along the ladder legs.

\begin{figure}[tb]
\centering
\includegraphics[width = 1\linewidth]{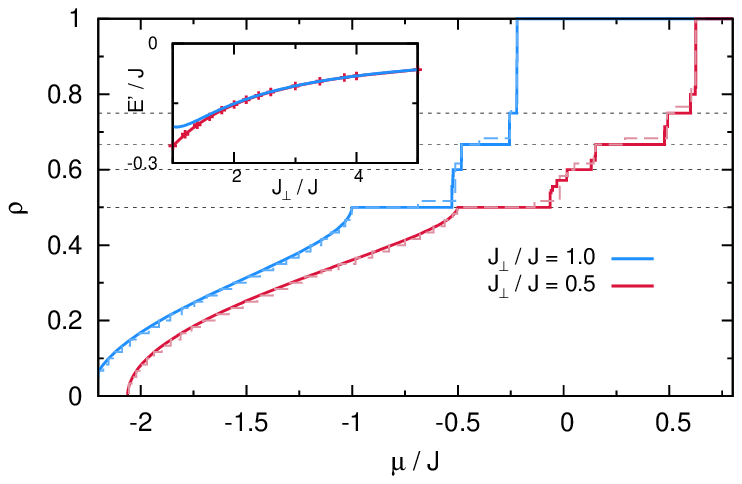}
\caption{Equation of state $\rho=\rho(\mu)$ for the $M=2$ for $U=\infty$, $\phi=\pi$ as obtained from finite system size DMRG and iDMRG calculations. Dashed lines correspond to DMRG computations with $L=60$ rungs, solid line is iDMRG data ($D=100$) for $\rho\geq 0.5$. The inset shows the ground-state energy $E' = e_0+\rho J_\perp$ of the iDMRG-calculation ($+$ symbols) for a filling $\rho=2/3$ as function of $J_\perp/J$ compared to perturbative result $-J/(3J_\perp) + J^4/(8 J_\perp^3)$ (solid line).}
\label{fig:comp_eqs_m2}
\end{figure}

In order to reduce finite-system size effects for a flux $\phi=\pi$ and $\rho>1/2$ we also perform iDMRG calculations, which turn out to be very efficient since convergence is reached for a small bond dimension $D<100$. We keep $p$ particles in $q$-site unit-cells with up to $q\leq 20$. The equation of state can then be approximated accurately from ${\rm min}_\rho(\{e_0(\rho) - \rho \mu\})$ with $\rho=p/q$ and $e_0(\rho)$ is the ground-state energy per ladder site.
This approach compares well to finite-system size DMRG-calculations with open boundary conditions as shown in Fig.~\ref{fig:comp_eqs_m2}. We also compare the iDMRG-calculations with results obtained from higher order perturbation theory for small values of $J/J_\perp$ (compare inset of Fig.~\ref{fig:comp_eqs_m2}).

The weak coupling approximation of Josephson junctions ladders~\cite{Kardar1986, Granato1990, Mazo1995, Denniston1995} may be described by a frustrated XY-model of classical spins
\begin{align}
\mathcal{H} \to -2J\rho \sum_{\zeta =0,1;j=1}^L  \cos( \theta_{\zeta,j+1} -\theta_{\zeta,j}) \nonumber\\
-2J_{\bot} \rho \sum_{j=1}^L \cos( \theta_{0,j}-\theta_{1,j} -j\phi).
\label{eq:H_JJ_1}
\end{align}
Assuming periodic boundary conditions and due to the property $\theta_{0,j+1} -\theta_{0,j} = \theta_{1,j} - \theta_{1,j+1}$, which is the case for the lowest energy configurations, Model~\eqref{eq:H_JJ_1} may be mapped to a one dimensional chain 
\begin{equation}
\mathcal{H} = \!-4J\rho\! \sum_{j=1}^L \left[  \cos(\frac{\alpha_{j-1}-\alpha_j}{2} + \phi) -\frac{J_{\bot}}{2J} \cos  {\alpha_j }\right],
\label{eq:H_JJ_2}
\end{equation}
with $\alpha_j=\theta_{0,j}-\theta_{1,j} -j\phi$.
The vortex-density dependence on flux, which we obtained in the ground state of classical one-dimensional phase-only Model~(\ref{eq:H_JJ_2}) by the effective potentials method~\cite{Griffiths1986,Mazo1995} is shown in Fig.~(2) of the main text.

\begin{table}[tb]
\begin{tabular}{cccl}
\hline
\hline
 \hspace{0.4cm}$\rho_V$, $\rho_H$\hspace{0.4cm} & \hspace{0.4cm}degeneracy\hspace{0.4cm} & configuration\\[0.1cm]
\hline
 $1/2$ & 2 & $\cdots \bullet \circ \bullet \circ \bullet \circ \bullet \circ \cdots$\\
\hline
 $2/5$ & 5 & $\cdots \circ \bullet \bullet \circ \bullet \circ \bullet \bullet \circ \bullet \cdots$ \\
\hline
 $1/3$ & 3 & $\cdots \circ \bullet \bullet \circ \bullet \bullet \circ \bullet \bullet \circ  \cdots$ \\
\hline
 $1/4$ & 4 & $\cdots \bullet \circ \bullet \bullet \bullet \circ \bullet \bullet \bullet \circ  \cdots$ \\
\hline
 $0$ & 1 & $\cdots\bullet \bullet  \bullet \bullet \bullet \bullet \bullet \bullet \cdots$\\
\hline
\hline
\end{tabular}
\caption{Ground-state degeneracies of some most stable VL/CDW states for ladders with periodic boundary conditions. Open circle indicates either a hole localized on a rung or a vortex localized in a plaquette. 
}
\label{tab:classical}
\end{table}

Ground-state degeneracies (for ladders with periodic boundary conditions) of some of the most stable VL/CDW configurations are listed in Table~SI.

\subsection{Stability of CDW states for flux $\phi<\pi$}

\begin{figure}[tb]
\centering
\includegraphics[width=0.45\columnwidth]{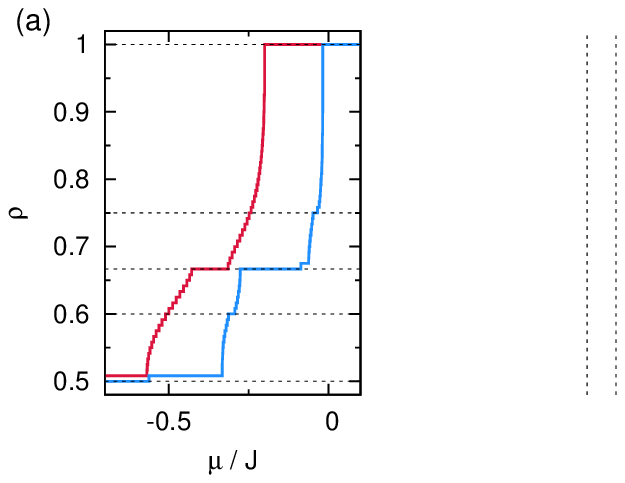}
\includegraphics[width=0.45\columnwidth]{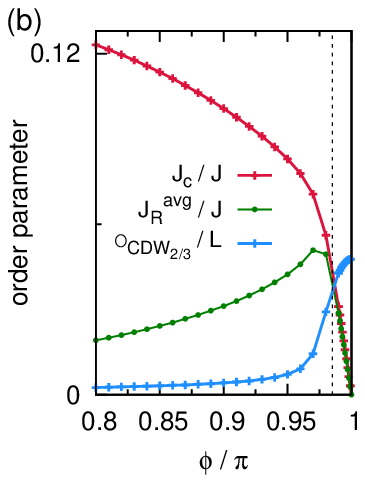}
\caption{Equation of state $\rho=\rho(\mu)$ for RHC ladder $J_\perp=J$ and
$\phi=0.99\pi$ and $\phi=0.999\pi$ (DMRG-calculation, $L=120$ rungs). The second curve has been shifted to the right for clarity. Note that due to the finite system size the curve consists of a series of finite steps. Only the large plateaus for $\phi=0.999\pi$ at $\rho=\frac{1}{2}$, $\frac{3}{5}$, $\frac{2}{3}$ and $\frac{3}{4}$ can be identified as CDW-phases. For $\phi=0.99\pi$ only plateaus at $\rho=\frac{1}{2}$ and $\frac{2}{3}$ remain.
(b) Chiral current $J_c/2$, average absolute rung-current $J_R^{avg}$ and density-structure factor $\mathcal{O}_{CDW_{2/3}}$ as function of the flux $\phi$ for filling $\rho=2/3$.
}
\label{fig:finitePhi_n2_3}
\end{figure}

Quantum gas experiments may allow for a smooth detuning of the magnetic flux to arbitrary values $0\leq \phi\leq \pi$ for which Model~(1) does no longer correspond to a  Hubbard model with two conserved Abelian quantum numbers. Hence, the fractional CDW-phases are soon destabilized as we depart from $\phi=\pi$ as shown in the $\mu-\rho$-curve of Fig.~\ref{fig:finitePhi_n2_3}~(a).

In Fig.~\ref{fig:finitePhi_n2_3}~(b) we plot the chiral current 
\begin{align}
J_c = -\frac{1}{L} \sum_{j,\gamma} (-1)^\gamma \mathcal{J}(j,\gamma\to j+1,\gamma) \,,
\end{align}
and the average absolute rung-current 
\begin{align}
J_R^{avg} = \frac{1}{L} \sum_j \left| \mathcal{J}(j,0\to j,1) \right| \,.
\end{align}
and the local particle current $\mathcal{J}(j,\gamma\to j',\gamma')$ from site $(j,\gamma)$ to site $(j',\gamma')$ is defined by the continuity equation $\langle \partial n_{(j,\gamma)} / \partial t \rangle = \sum_{\langle (j',\gamma')\rangle} \mathcal{J}(j,\gamma\to j',\gamma')$.
For the case of the CDW$_{2/3}$ phase as $\phi\approx\pi$ the chiral and average rung-current increase linearly with the same slope when reducing flux from $\phi=\pi$. The CDW-order parameter $\mathcal{O}_{CDW_{2/3}}=S(k=2\pi/3)$ strongly decreases in M-SF phase.

\begin{figure}[tb]
\centering
\includegraphics[width=0.45\columnwidth]{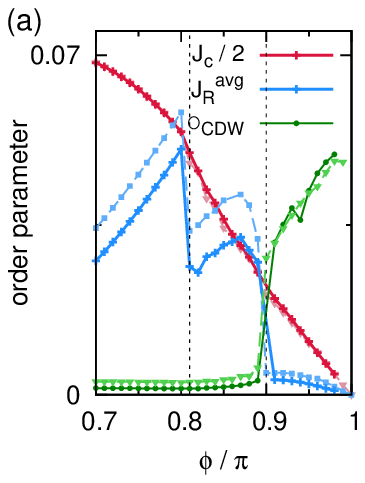}
\includegraphics[width=0.45\columnwidth]{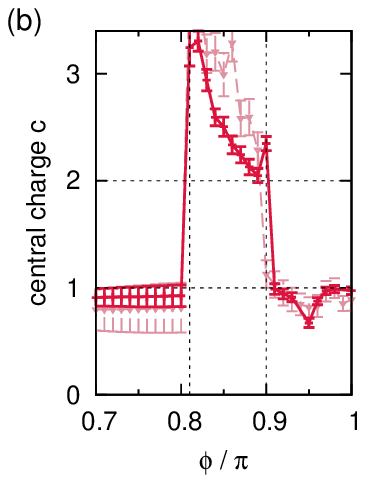}
\caption{Behavior of currents and of central charge along the cut through the phase diagram, presented in Fig.~4 of the main text, when decreasing flux $\phi$ from $\pi$ for $\rho=0.4$. 
(a) Chiral current $J_c/2$, average absolute rung-current $j_R$ and density-structure factor $\mathcal{O}_{CDW}=S(k=\pi/2)$ for $L=61$ (plus) and $L=121$ (boxes) rungs.
(b) Estimated central charge $c$ for $L=61$ (triangles) and $L=121$ (plus) rungs from fit of the block-entanglement entropy to the Calabrese-Cardy formula~\cite{Holzhey1994, Vidal2003, Korepin2004, Calabrese2004}. Due to the finite-size effects the central charge is over(under) estimated.}
\label{fig:finitePhi_ss}
\end{figure}

Fig.~\ref{fig:finitePhi_ss} shows a detailed cut through phase diagram of Fig. 4 presented in the main text at a low filling $\rho=0.3$. The V-SF to the SS transition can be seen in a sharp jump in central charge (see Fig.~\ref{fig:finitePhi_ss}~(b)) and an increase of charge density order (see Fig.~\ref{fig:finitePhi_ss}~(a)), which we for the case of open boundary DMRG calculations define by the peak of the static structure factor $\mathcal{O}_{CDW}=S(k=2 \pi\rho)$ with
\begin{align}
S(k)=\frac{1}{L} \sum_k \e^{i k (i-j)} \langle n_i n_j \rangle \,.
\end{align}

\begin{figure}[tb]
\centering
\includegraphics[width=1\columnwidth]{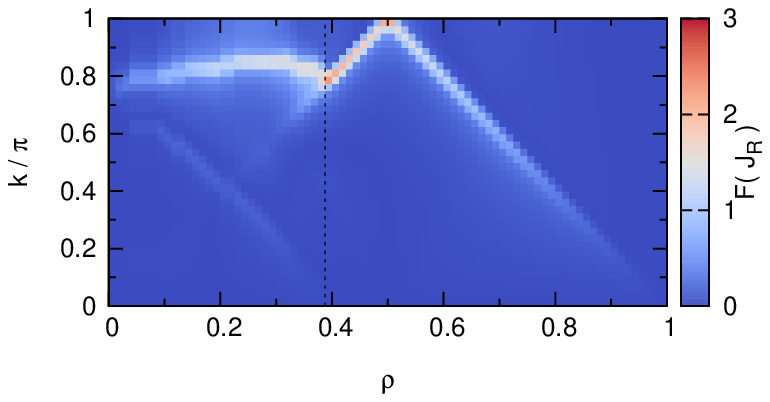}
\caption{Fourier transform of the real-space pattern of the rung-currents $F(J_R)(k)$ for a cut through phase diagram Fig.~(4) of the main text as function of the filling $\rho$ for $\phi=0.8\pi$. The dashed line denotes the estimated phase transition between V-SF ($c=2$) to the Laughlin or Meissner-like phases with central charge $c=1$ (DMRG-calculation, $L=80$, $U=\infty$, $J_\perp=J$).}
\label{fig:ccfft_phi0.8}
\end{figure}

In Fig.~\ref{fig:ccfft_phi0.8} we show the Fourier transform $F(J_R)$ of the real-space patterns of the rung currents $\langle J_R \rangle(j)$ for a cut through the phase diagram of Fig.~(4) of the main text. At the transition from the V-SF to the M-SF-like phase (which we also identify by observation of a pronounced kink in the $\mu-\rho$-curve and a change in central charge, see e.g. Ref.~\cite{Piraud2015} for details) we observe a drastic change in the behavior of $F(J_R)$.
Generally, close to the V-SF boundary for $\phi\gtrsim 0.6$ we observe a small region of enhanced (boundary driven) rung-current fluctuations (which are accompanied by local density oscillations). This kind of incommensurate Meissner-phases may be adiabatically connected to the commensurate M-SF phase observed for $\phi\to 0$~\cite{Greschner2016} and, recently, have been connected to a certain class of Laughlin-precursor states in Ref.~\cite{Petrescu2016}. For finite rung-interactions different Laughlin-like states have been also discussed in Ref.~\cite{Strinati2016}.

\subsection{Stability of CDW states for finite interactions $U<\infty$}

In the following we discuss the stability of the CDW-phases for a finite interaction $U<\infty$ of the Model~(1) presented in the main text.

\begin{figure}[tb]
\centering
\includegraphics[width=1\columnwidth]{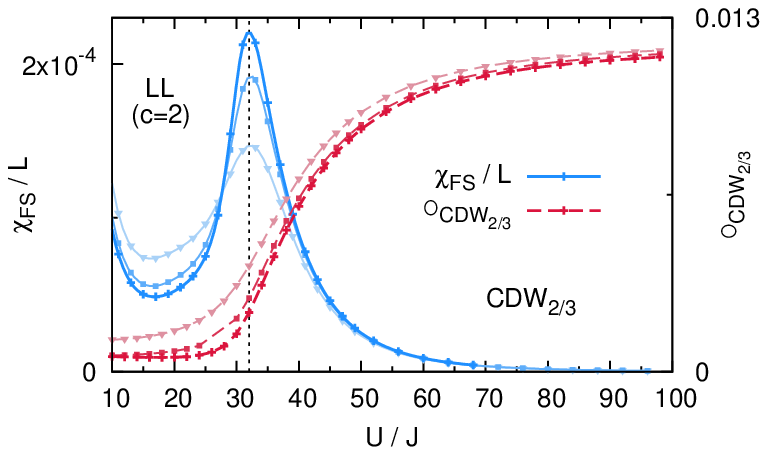}
\caption{Stability of the CDW$_{2/3}$-phase for $M=2$ fermions as functions of $U / J$. Shown are the fidelity susceptibility $\chi_{FS} / L$ and the CDW-order parameter $\mathcal{O}_{CDW_{2/3}}$ for system sizes $L=60$ (triangles), $L=120$ (boxes) and $L=180$ (plus).}
\label{fig:finiteU_F}
\end{figure}

As shown in Fig.~\ref{fig:finiteU_F} for fermions for finite $U$ we find a transition to a gapless state with a central charge $c=2$ for $J_\perp=J$ around $U\approx 30J$ signaled by strong decrease in the CDW-order parameter $\mathcal{O}_{CDW_{2/3}}$ and a slowly increasing peak in the fidelity susceptibility
\begin{align}
\chi_{FS}(U) = \lim_{\delta U\to 0} \frac{-2 \ln |\langle \Psi_0(U) | \Psi_0(U + \delta U) \rangle| }{(\delta U)^2} \,,
\end{align}
with $|\Psi_0\rangle$ being the ground-state wavefunction.
The scaling of the peak ${\rm max} \chi_{FS}(\phi)$ with the system size is consistent with a logarithmic convergence as expected for a BKT-type transitions~\cite{Sun2015,Vekua2016}. 

\begin{figure}[tb]
\centering
\includegraphics[width=1\columnwidth]{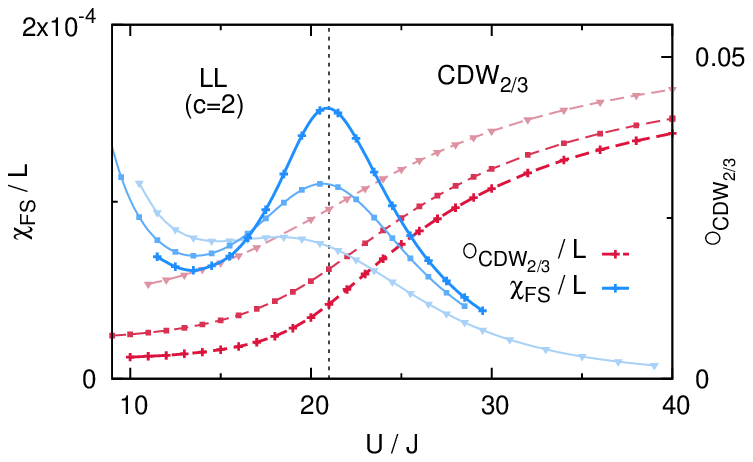}
\caption{Stability of the CDW$_{2/3}$-phase for $M=2$ boson as functions of $U / J$. Shown are the fidelity susceptibility $\chi_{FS} / L$ and the CDW-order parameter $\mathcal{O}_{CDW_{2/3}}$ for system sizes $L=30$ (triangles), $L=60$ (boxes) and $L=120$ (plus). We employ a cutoff to the bosonic local Hilbert-space of $n_{max}\leq 6$ bosons per site.}
\label{fig:finiteU_B}
\end{figure}

Interestingly, the CDW-phases stay more stable for bosons with reducing $U$, as shown in Fig.~\ref{fig:finiteU_B} for the CDW$_{2/3}$ phase. The stability of the supersolid phase for bosonic atoms has been studied in Ref.~\cite{Bilitewski2016} employing an effective model valid in the strong-rung coupling limit.

\subsection{Connection to the thin-cylinder limit of quantum Hall system}

For $M>2$ we start with $SU(M)$ Hubbard-model with a uniform flux and put periodic boundary conditions along the leg and rung directions, to make contact with quantum Hall system in thin torus limit,
\begin{align}
H =& - J \sum_{j=1, \zeta}^L [b_{j+1,\zeta}^\dagger b_{j,\zeta} +b^\dagger_{j,\zeta}b_{j+1,\zeta}]-\mu\sum_{j}n_j \\ 
&-  J_\perp \sum_{j,\zeta } [ \e^{\ii \phi j} b_{j,\zeta}^\dagger b_{j,\zeta+1} + \rm{H.c.}] + \frac{U}{2} \sum_{j,\zeta,\zeta'} n_{j,\zeta} n_{j,\zeta'}\nonumber,
\label{eq:SD_orig_rung}
\end{align}
where summation is for $\zeta=0,1,...,M-1$.
The number of sites along the real dimension $L$ will be assumed much larger than $M$.

\begin{figure}[tb]
\centering
\includegraphics[width = 0.32\linewidth]{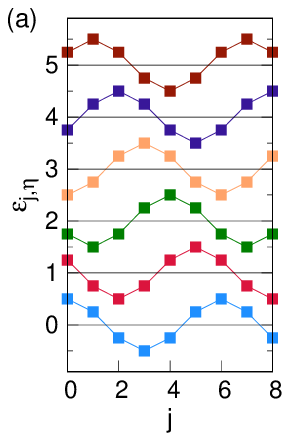}
\includegraphics[width = 0.32\linewidth]{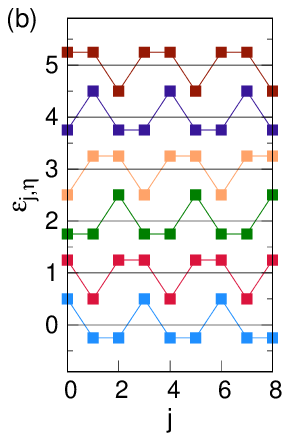}
\includegraphics[width = 0.32\linewidth]{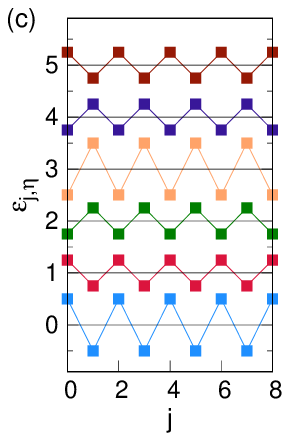}
\caption{Examples of $\epsilon_{j,\eta}=\cos{\frac{2\pi(\eta+mj)}{M}}$ for $M=6$ and (a) $m=1$, hence $\phi=\frac{\pi}{3}$, (b) $m=2$, $\phi=\frac{2\pi}{3}$ and (c) $m=3$, $\phi=\pi$ (curves have been shifted for clarity).}
\label{fig:epsilonxy}
\end{figure}

Values of fluxes that are compatible with the gauge invariance of the model on cylinder geometry are $\phi M=2m \pi $ (where $m$ is an integer dividing $M$). We introduce the new ``flavor'' basis ~\cite{Barbarino2015}, which conserves exactly $M$ particle numbers separately $\sum_j  \tilde n_{j,\eta}$, $\eta=0,1,..M-1$ as it is the case for $\phi=\pi$ in the ladder case. We perform a canonical transformation of the above Hamiltonan to the flavor basis, given by the Fourier transformation along the synthetic direction:
\begin{equation}
b_{j,\zeta}=\frac{1}{\sqrt{M}}\sum_{\eta=0,.M-1} e^{i\zeta \eta 2\pi/M} \tilde b_{j,\eta},
\end{equation}
where
\begin{equation}
\tilde b_{j,\eta}=\frac{1}{\sqrt{M}}\sum_{\zeta=0,..M-1} e^{-i\zeta \eta 2\pi/M}  b_{j,\zeta}.
\end{equation}
Then, in this flavor basis the model on the $M$-leg cylinder is given by
\begin{align}
H=& -J \sum_{j,\eta}[ \tilde b_{j,\eta}^\dagger \tilde b_{j+1,\eta} + H.c.]  - J_\perp \sum_j \epsilon_{j,\eta} \tilde n_{j,\eta}\\
&+ \frac{U}{2} \sum_{j} n_{j} (n_{j}-1)-\mu\sum_{j}n_j, \nonumber
\end{align}
where $\eta$ is a flavor index $\eta=0,1,..M-1$ and local flavor dependent chemical potentials are given by $\epsilon_{j,\eta} =  \cos\left(\frac{2\pi (\eta+mj)}{M}  \right)$.

At $U=\infty$ and $J_\perp=0$ there is an extensive degeneracy with respect to flavor index $\eta$ of any state, resulting in a flat flavor band analogous of flat spin band for $M=2$ case. $J_\perp$ lifts this degeneracy due to $\epsilon_{j,\eta}$ dependence on $\eta$. Examples for $\epsilon_{j,\eta}$ for $M=6$ are shown in Fig.~\ref{fig:epsilonxy} for $m=1,2$ and $3$. Note, that at $\phi=\pi$  $\epsilon_{j,\eta}$ selects only two flavor components in the ground state, $\eta=0$ and $\eta=3$. 
One can see that at $\pi$-flux the physics is similar to $M=2$ case, and namely for low filling $\rho<1/2$ there is a supersolid state and CDW staircase for $1/2<\rho<1$. We remind that case corresponding to one particle at each rung has $\rho=1$ in our convention. 
This is generically so for any even $M$ at $\pi$- flux. Similarly, at $\phi=2\pi/3$ only three components are important, corresponding to odd $\eta$, for which $\epsilon_{j,\eta}=1$ for some values of $j$. This is also true for any $M$ multiple of $3$. At $\phi=\pi/6$ all $6$ flavour components are equally important. Extrapolating analogous considerations to $M\to \infty$ results in the phase diagram presented in the inset of Fig. 4 in the main text.

\begin{figure}[tb]
\centering
\includegraphics[width = 1\linewidth]{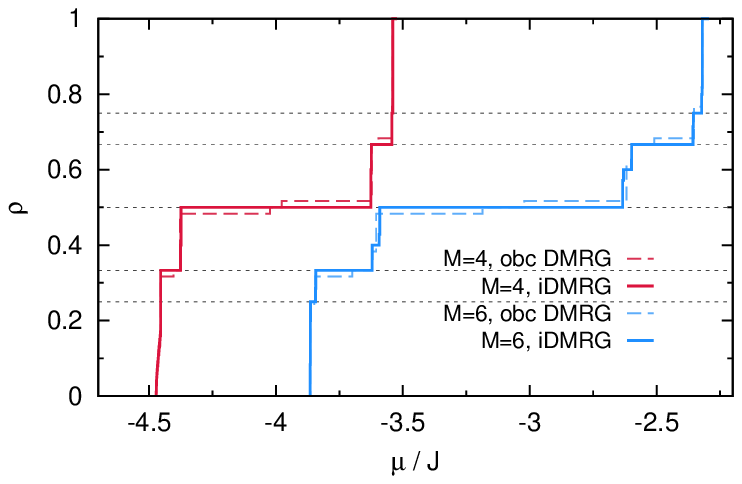}
\caption{Equation of state $\rho=\rho(\mu)$ for the cylinder-geometry for (a) $M=4$ and (b) $M=6$ legs ($J_\perp=2J$, $U=\infty$, shifted for clarity). Dashed lines correspond to DMRG computations with $L=60$ rungs, solid line is iDMRG data for $\rho>\phi/2\pi$, for $\phi=\frac{2\pi}{M}$.}
\label{fig:torus_eqs}
\end{figure}

\begin{figure}[tb]
\centering
\includegraphics[width = 0.48\linewidth]{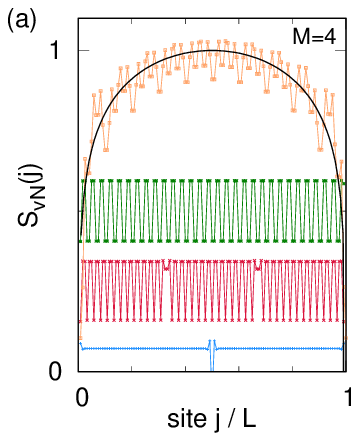}
\includegraphics[width = 0.48\linewidth]{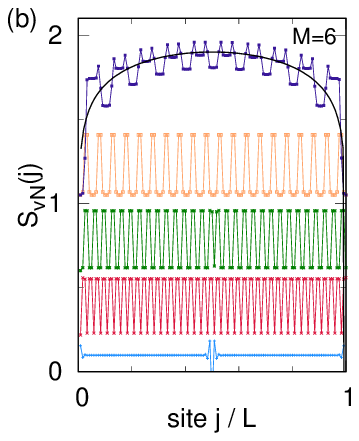}
\caption{Examples of the block entanglement entropy for various densities of the $M$-leg cylinder for $\phi=\frac{2\pi}{M}$ (DMRG data, $L=122$, $J_\perp=2J$, $U=\infty$, curves have been shifted for clarity). (a) $M=4$ legs and (from top to bottom) $\rho=1/6$, $1/3$, $1/4$, $1/2$ (b) $M=6$ legs and (from top to bottom) $\rho=1/12$, $1/6$, $1/4$, $1/3$, $1/2$. The dashed line is fit to Calabrese-Cardy-formula~\cite{Holzhey1994, Vidal2003, Korepin2004, Calabrese2004} for $c=1$.}
\label{fig:torus_ee}
\end{figure}

To further verify these assumptions we present DMRG and iDMRG calculations for tori of $M=4$ and $M=6$ rungs at $\phi=\frac{2\pi}{M}$, in order to obtain the $\mu-\rho$-curves (see Fig.~\ref{fig:torus_eqs}). In Fig.~\ref{fig:torus_ee} the von-Neumann entanglement entropy $S_{vN}(j)$ is shown for several densities $\rho$, which corresponds to gapless state with a central charge $c=1$ for $\rho<\phi/2\pi$~\cite{Holzhey1994, Vidal2003, Korepin2004, Calabrese2004}, and gapped CDW-states at higher fillings.

For the continuous two-dimensional quantum Hall system, filling $\nu$ is defined as a ratio of the total number of electrons to the total number of fluxes. For the cylinder geometry lattice model we analogously can define
\begin{equation}
\nu=\frac{N}{N_\phi}=\frac{N}{M (L-1)\phi/2\pi}=\frac{2\pi \rho}{M\phi},
\end{equation}
where one-dimensional density is $\rho=\frac{N}{L}$.
Choosing $\phi=\frac{2\pi m}{M}$ gives
\begin{equation}
\label{nurho}
\nu=\frac{\rho}{m}.
\end{equation}
This gives that for $M=3$ (where only one value $m=1$ is possible) filling $\rho=1/3$ corresponds to $\nu=1/3$ state, whereas for $M=6$ both $\rho=1/3$, $m=2$ and  $\rho=1/6$, $m=1$ correspond to the $\nu=1/3$ state.

From Eq.~(\ref{nurho}) for $m=1$ it follows that CDW states for $M$-leg cylinder at $2\pi/M$-flux realized for $1/M\le\rho\le 1$ relate to quantum Hall states in thin-cylinder limit corresponding to filling factor $ 1/M \le \nu\le 1 $. However, only for sufficiently large $M$ it makes sense to address topological properties of $M$-leg ladder in resemblance of fractional quantum Hall states~\cite{Zeng2015,Saito2016}.

\bibliography{references}